\pdfoutput=1
\documentclass[prl,twocolumn,showpacs,floatfix]{revtex4-1}
\usepackage{times,amsmath,amssymb,amstext,latexsym,float,graphicx,color}
\usepackage{hyperref}
\hypersetup{colorlinks=true, citecolor=blue, urlcolor=blue, linkcolor=blue}

\begin{document}

\title{Rotating binary Bose-Einstein condensates and vortex clusters in quantum droplets}

\author{M.~Nilsson~Tengstrand$^{1}$, P. St\"urmer$^{1}$, E. Karabulut$^{2}$ and S.~M.~Reimann$^{1}$}
\affiliation{$^1$Mathematical Physics and NanoLund, Lund University, Box 118, 22100 Lund, Sweden\\
$^2$Department of Physics, Faculty of Science, Selcuk University, TR-42075 Konya, Turkey}

\date{\today}

\begin{abstract}
Quantum droplets may form out of a gaseous Bose-Einstein condensate, stabilized by quantum fluctuations beyond mean field. We show that multiple singly-quantized vortices may form in these droplets at moderate angular momenta in two dimensions. Droplets carrying these precursors of an Abrikosov lattice remain self-bound for certain timescales after switching off an initial harmonic confinement. Furthermore, we examine how these vortex-carrying droplets can be formed in a more pertubation-resistant setting, by starting from a rotating binary Bose-Einstein condensate and inducing a metastable persistent current via a non-monotonic trapping potential.
\end{abstract}

\maketitle

The formation of self-bound droplets is a well-known macroscopic phenomenon. For an exemplary droplet of water, stability and shape rely on the balance of effective forces between its constituent particles -- attractive ones that keep it together, and repulsive ones that prevent it from collapse. Their interplay defines the droplets' surface tension, stabilizing the system in a metastable state. 
Such droplets do not only occur at a macroscopic level, but are ubiquitous also in the quantum realm, where nuclei~\cite{BohrMottelson} and superfluid helium droplets~\cite{donelly1991,toennies2004,ancilotto2018} are prominent examples. While these are rather dense and strongly interacting many-body systems, recent experiments with ultra-cold quantum gases of bosonic atoms uncovered a novel type of quantum liquid: 
Self-bound droplets may form out of a gaseous Bose-Einstein condensate (BEC) of dysprosium~\cite{kadau2016,schmitt2016,ferrierbarbut2016,ferrierbarbut2016b,ferrierbarbut2018} or erbium~\cite{chomaz2016}, atomic species that are known for their strong dipolar interactions~\cite{lu2010,lu2011,aikawa2012}. Similar droplet states  have more recently also been realized with binary Bose gases of potassium in different hyperfine states~\cite{cabrera2018,semeghini2018}, 
where the inter- and intracomponent interactions are short-ranged. These quantum droplets can be large, containing thousands of atoms. Importantly, they are very dilute -- by more than eight orders of magnitude when compared with liquid helium~\cite{cabrera2018}. While the discovery with dysprosium~\cite{kadau2016,schmitt2016,ferrierbarbut2016} at first came as a surprise, the binary self-bound droplet states were theoretically predicted a year before~\cite{petrov2015} for a scenario similar to the experiments with potassium, and also in lower dimensions~\cite{petrov2016}. That higher-order corrections beyond mean field may lead to self-bound states was discussed earlier in a different setting in Refs.~\cite{bulgac2002,hammer2004}.
For the dipolar or binary self-bound bosonic systems of Refs.~\cite{kadau2016,schmitt2016,ferrierbarbut2016,semeghini2018,cabrera2018} 
the physical mechanism of droplet formation is based on tuning the interactions in gas such that only a weak effective attraction remains. While in pure mean field this would lead to a collapse of the system, weak first-order corrections to the mean field energy, often referred to as the Lee-Huang-Yang (LHY)-correction~\cite{lhy1957}, can become comparable in size and may thus stabilize the system.

Bound states that are  merely a consequence of quantum corrections beyond mean field have been known since long from alkali-metallic clusters~\cite{knight1984,nishioka1990}, where the jellium of the ionic charge background cancels the electronic Hartree term, resulting in a liquid-like electronic state bound mainly due to a balance between kinetic and exchange-correlation energy~\cite{koskinen1995}. The fact that self-bound bosonic droplets can form with atoms as intrinsically different as lanthanides and alkali metals shows that this phenomenon for ultra-cold gases is a general one, giving evidence for a novel state of quantum matter with new and unexpected properties.

In atomic quantum droplets the effective interactions between the constituent bosonic atoms are relatively weak. This eases their theoretical description, making it largely accessible to lowest-order corrections beyond mean field~\cite{petrov2015}. On the theoretical side, progress has been made in the framework of both the extended Gross-Pitaevskii approach~\cite{baillie2016,wachtler2016,wachtler2016b,chomaz2016,bisset2016,baillie2016,baillie2017} where the LHY-correction as well as atom losses are added to the non-linear Schr\"odinger equation in an efficient ad hoc manner, 
by quantum Monte-Carlo approaches~\cite{saito2016,macia2016,cinti2017,cikojevic2018},  
or by solving the Bogoliubov-de Gennes equations~\cite{bisset2016,baillie2017}. 

As these droplets form out of a BEC there is good reason to assume that they
have  superfluid properties. One of the signatures of superfluidity are
vortices -- topologically non-trivial states well known from harmonically
trapped  BECs~(see e.g.~\cite{butts1999,matthews1999,
madison2000,chevy2000,kavoulakis2000,aboshaeer2001,raman2001,madison2001,haljan2001}
or the reviews~\cite{fetter2009,saarikoski2010}),
characterized by a depletion of the density accompanied by a phase shift.
So far, however, experimental evidence for vortices in these droplets
appears  elusive. Only a few theoretical works yet considered the
LHY-stabilized quantum droplets' rotational properties. Very recent work
found metastable necklace-like clustered droplets carrying angular
momentum~\cite{kartashov2019}.
Stability of an imprinted singly quantized vortex was reported for a
prolate dipolar droplet~\cite{cidrim2018}, accompanied by a  splitting of
the droplet into two smaller droplet fragments. Imprinted vortices at the
droplet center with similar fragment formation were also reported for
binary droplets in both two~\cite{li2018} and three dimensions, in the latter case 
carrying up to two units of angular momentum in a region of experimentally
accessible parameters~\cite{kartashov2018}. A doubly quantized vortex was
found to decay into two singly-quantized vortices upon a quadrupolar
deformation~\cite{ancilotto2018b}. In these studies, the vorticity was imprinted on the droplet by a phase factor of the initial state.

In this Letter we investigate the rotational properties of a trapped binary BEC in relation to self-bound quantum droplets. We find that by starting from the rotational ground state of this binary BEC, metastable droplets containing vortex clusters may form after a sufficiently slow release from a harmonic confinement. We also explore the feasibility of creating vortex droplets in this way by starting from the ground state of a trapped condensate with a non-monotonic trapping potential, where we find the existence of a metastable persistent current.

Let us now consider a species-symmetric binary BEC confined to two dimensions~\cite{petrov2016} that is interacting weakly via short-range interactions, where the inter- and intraspecies
interactions are assumed to be attractive and repulsive, respectively. 
For such a binary BEC with equal masses of the atoms in the two components, the coupled Gross-Pitaevskii equations reduce to that of a one-component BEC with an accordingly modified interaction term ~\cite{petrov2015,petrov2016}.  We initally confine the gas in a harmonic trap with an added Gaussian at the trap center. The LHY-amended Gross-Pitaevskii equation for such a system in a frame rotating with angular frequency $\Omega$ can then be written as
\begin{eqnarray}\label{gp}
i\frac{\partial\psi}{\partial t} = \Big(-\frac{1}{2}\nabla^2 +
\frac{1}{2}\omega^2{r}^2 + V_0e^{-(r/a_{\perp})^2} + \nonumber \\ 
+~\left|\psi\right|^2\ln\left|\psi\right|^2  - iL_3\left|\psi\right|^4 -
\Omega L_{z}\Big)\psi,
\end{eqnarray}
\noindent where the scaling invariances of the system have been used to
bring the equation into this dimensionless form. Here $\omega$ is the harmonic trapping
frequency, $V_0$ the amplitude of the Gaussian,
$a_{\perp}=1/\sqrt{\omega}$ the oscillator length and $L_3$ the rate of
three-body losses. The order parameter is normalized according to $\int
d^2\mathbf{r}|\psi|^2 = N$.  The energy in the non-rotating frame is 
\begin{align}\label{gpenergy_nonrot}
E = \int d^2\mathbf{r} \Bigg[ \frac{1}{2}|\nabla\psi|^2 &+
\frac{1}{2}\omega^2 r^2|\psi|^2 + V_0 e^{-(r/a_{\perp})^2}|\psi|^2 
+ \nonumber \\
~\qquad &+\frac{1}{2}|\psi|^4\ln{\left(\frac{|\psi|^2}{\sqrt{e}}\right)}\Bigg]
\end{align}
and the angular momentum $L = \int d^2\mathbf{r}\psi^\ast L_z \psi$. Equation~(\ref{gp}) is solved with the usual split-step Fourier method~\cite{chinkrotscheck2005} in real and imaginary time. For the imaginary time-propagation we use set of different randomly perturbed initial conditions in order to avoid local minima in the energy landscape. 

\begin{figure}
  \centering
  \includegraphics[width = 0.49\textwidth]{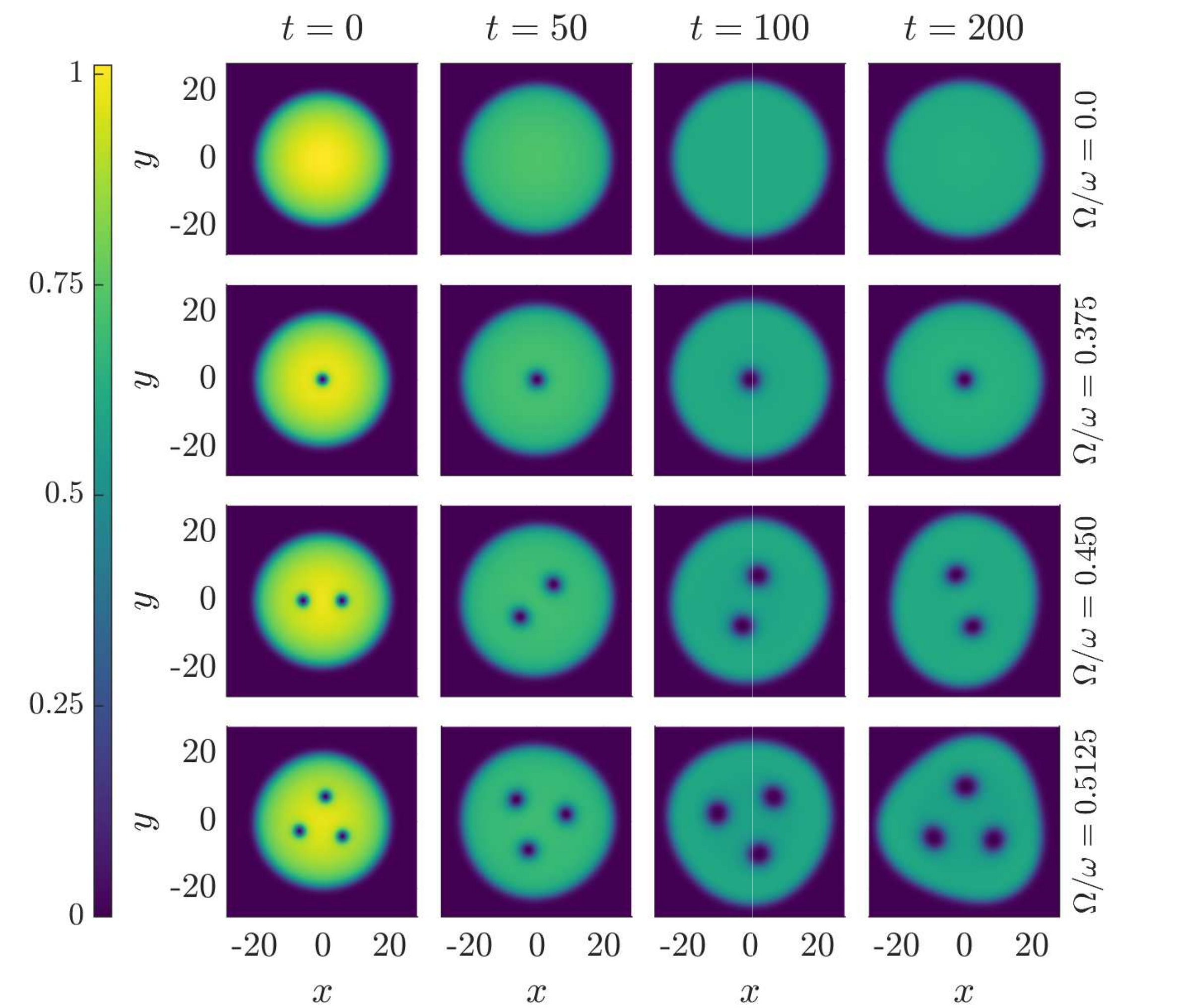}
  \caption{Time evolution of various rotational ground states in a harmonic 
  trap ($V_0 = 0$) at different rotation frequencies for $N=1000$, $\omega=0.04$ and $L_3 = 0$ (i.e. no three-body losses). 
  The harmonic trap frequency is linearly decreased in time such
that it is zero at $t = 100$.}
  \label{fig:panelsho}
\end{figure}

We first look for the ground state of free droplets in a rotating frame, but before convergence can be reached, the droplets are found to decay to fragments similar to those in the three-dimensional case \cite{kartashov2018}. As a first remedy to these inherently unstable solutions, we add a weak stabilizing harmonic confinement to the system. The dimensionless parameters considered here to illustrate our findings are
$N=1000$ and $\omega=0.04$. For these values the corresponding free
droplet has the characteristic flat-top shape \cite{li2018}, and
the trapping frequency is sufficiently weak to keep the droplet at a density close to its (free) equilibrium value. We first of all consider a purely harmonic trap ($V_0 = 0$) and identify the rotational ground states at
distinct rotation frequencies $\Omega$. The density distributions for some
ground states at different $\Omega$ are shown in the leftmost column of
Fig.~\ref{fig:panelsho}.
\begin{figure}
\centering
  \includegraphics[width = 0.49\textwidth]{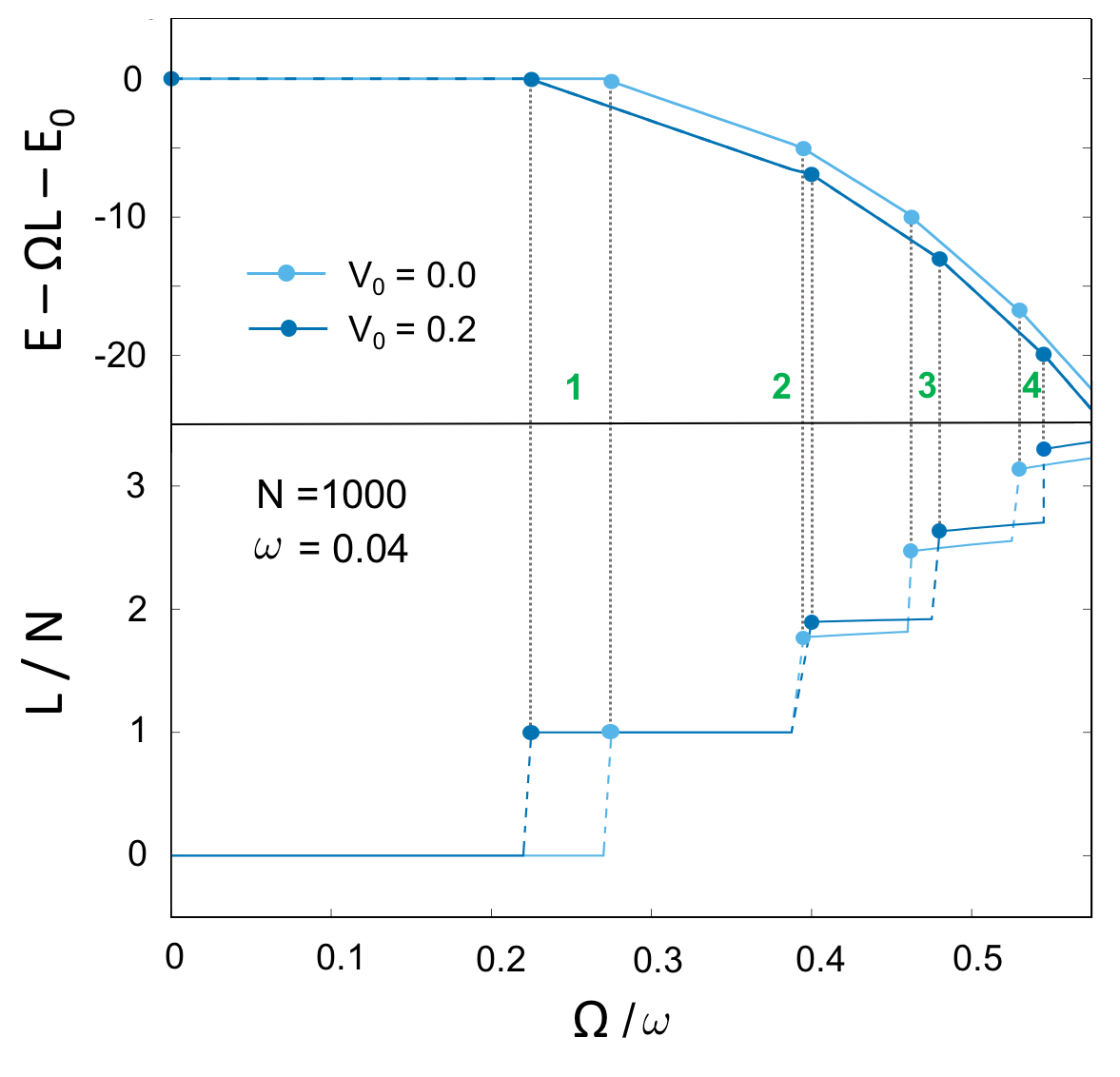}
  \caption{{\it Upper panel:} Ground state energy in the rotating frame as a function of rotation frequency  for $N=1000$ and $\omega =0.04$, in a harmonic trap ($V_0=0$, {\it light blue}) and with a Gaussian added at the center ($V_0=0.2$, {\it dark blue}). Both curves are plotted against the energy of their respective non-rotating ground state $E_0$. {\it Lower panel:} Angular momentum as a function of rotation frequency for the same parameters. Distinct kinks in the energy correspond to an increase in the number of singly quantized vortices {\it (green numbers)} displayed as steps in the angular momentum shown in the lower panel.}
  \label{fig:omega}
\end{figure}
Clearly, with increased rotation, vortices are
induced in a way similar to that of one-component condensate, proceeding
from the formation of a unit vortex at the trap center, to two- and
three-vortex states with the usual two- and three-fold symmetries. Figure~\ref{fig:omega} shows the ground state energy in the rotating frame $E-\Omega L$ and the corresponding angular momentum as a function of the trap rotation. The first three
steps in $L$, corresponding to kinks in the rotational energy, are
seen for $L/N \approx 1.0, 1.8$ and $2.5$ for the first three
vortex states shown in Fig.~\ref{fig:panelsho} similarly to scalar
BECs~\cite{butts1999,kavoulakis2000}.

\begin{figure}
\centering
  \includegraphics[width = 0.49\textwidth]{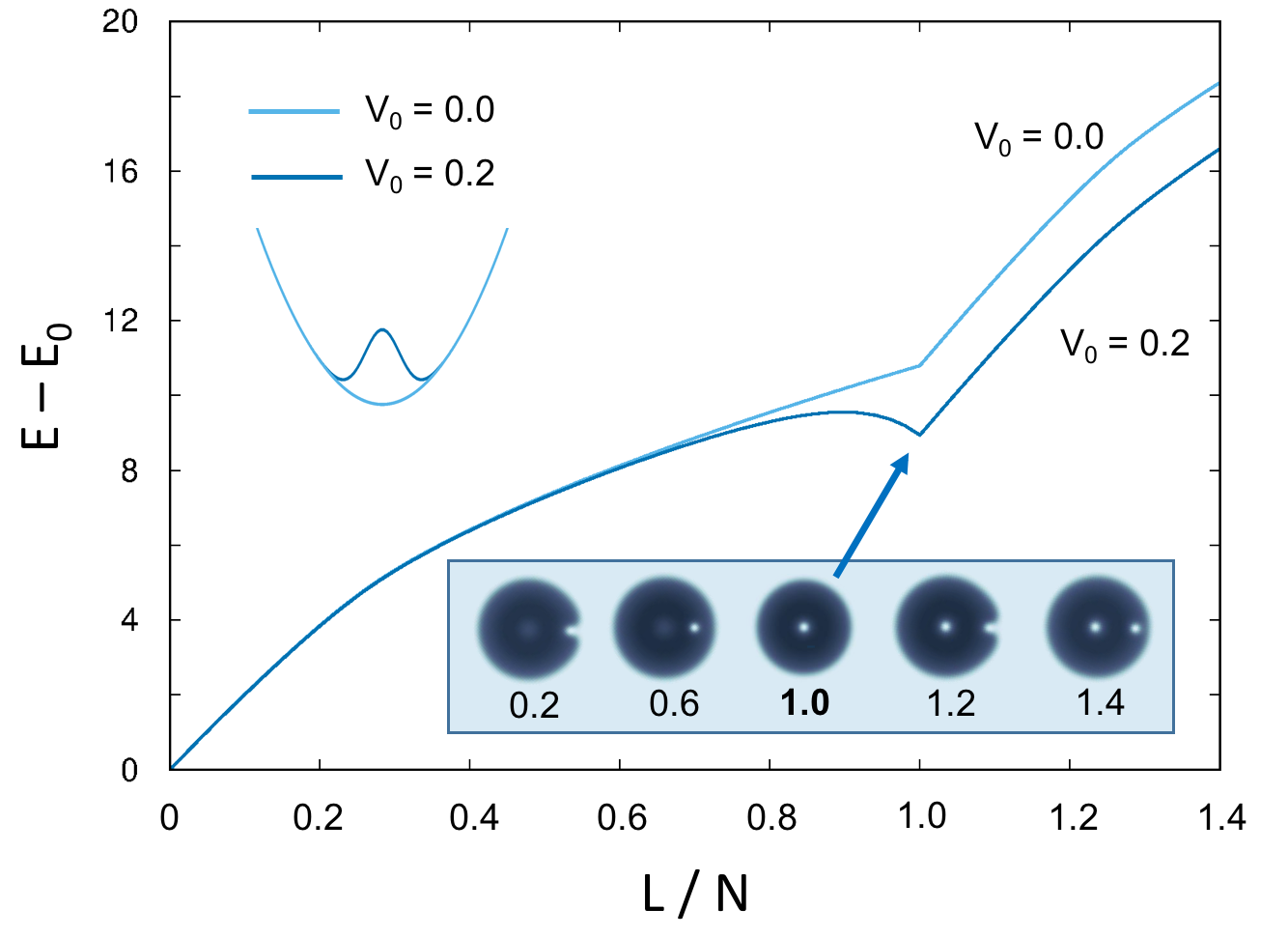}
  \caption{Energy as a function of angular momentum for $N=1000$ and $\omega =0.04$, in a harmonic trap ($V_0=0$, {\it light blue}) and with a Gaussian added at the trap center ($V_0=0.2$, {\it dark blue}), as in Fig.~\ref{fig:omega}. Both curves are plotted against the energy of their respective non-rotating ground state $E_0$.}
  \label{fig:dispersion}
\end{figure}

Instead of minimizing the energy by solving Eq.~(\ref{gp}) for
fixed $\Omega$, one may instead study the energy at a fixed angular
momentum by minimizing $E + C(L-L_0)^2$ \cite{komineas2005}, where $C$ and $L_0$ are dimensionless constants. For sufficiently large values
of $C$, the energy minimum then occurs at an angular momentum $L=L_0$,
making it possible to obtain solutions for states that are not rotational ground states, i.e. for arbitrary $L$. The dispersion relation obtained in this way is displayed in
Fig.~\ref{fig:dispersion} for angular momenta up to and beyond the unit vortex
that nucleates at the trap center at $L/N=1.0$. At this value, for
$V_0=0$, the energy has a kink, which can turn into a energetic minimum in
the rotating frame when the energy $E-\Omega L$ is tilted downwards by
a constant slope, resulting in the first step in $L$ when solving
Eq.~(\ref{gp}) for a corresponding value of $\Omega$, see 
Fig.~\ref{fig:omega}. For higher vortex numbers the mechanism is similar,
with kinks in $E$ leading to the plateaus in $L$ for certain
values of $\Omega$. In the absence of three-body losses, $L_3=0$, Eq.~(\ref{gp}) conserves angular momentum. This implies that a
condensate in a unit vortex ground state in the rotating frame will remain
so even after the rotation ceases by virtue of this conservation law.
However, since there is no local minimum in the dispersion relation for
the purely harmonic case, this state is susceptible to small pertubations. It will thus slide down in energy to the non-rotating ground state. Let us next consider a
non-monotonic trapping potential by adding a Gaussian to the center of the
harmonic trap (as has been realized experimentally, see for example~\cite{bretin2004}). Trapping potentials of this kind have been
shown to cause local minima in the dispersion relation for 
scalar BECs~\cite{karkkainen2007}. The energy as a function of angular
momentum for $V_0 = 0.2$ shown in Fig.~\ref{fig:dispersion} confirms that
this is also the case for this binary system. Such a mexican-hat type of confinement
can thus support a metastable persistent current even in the presence of
weak pertubations. The energy in the rotating frame and the corresponding
angular momentum as a function of rotation frequency with this central Gaussian is shown in Fig.~\ref{fig:omega}.

Since we are interested in rotational properties of self-bound
condensates, we now imagine a scenario where the ground state at a
particular rotation frequency is maintained even after the trap rotation
has stopped (as could be realistic in an experimental setting when there
exists a local minimum in the dispersion relation). The condensate is then
released from the trap by decreasing $\omega$ linearly in time in order to
reduce the radial velocity that results from the expansion. The real time propagation
for the purely harmonic case is shown in Fig.~\ref{fig:panelsho}. 
\begin{figure}
\centering
  \includegraphics[width = 0.49\textwidth]{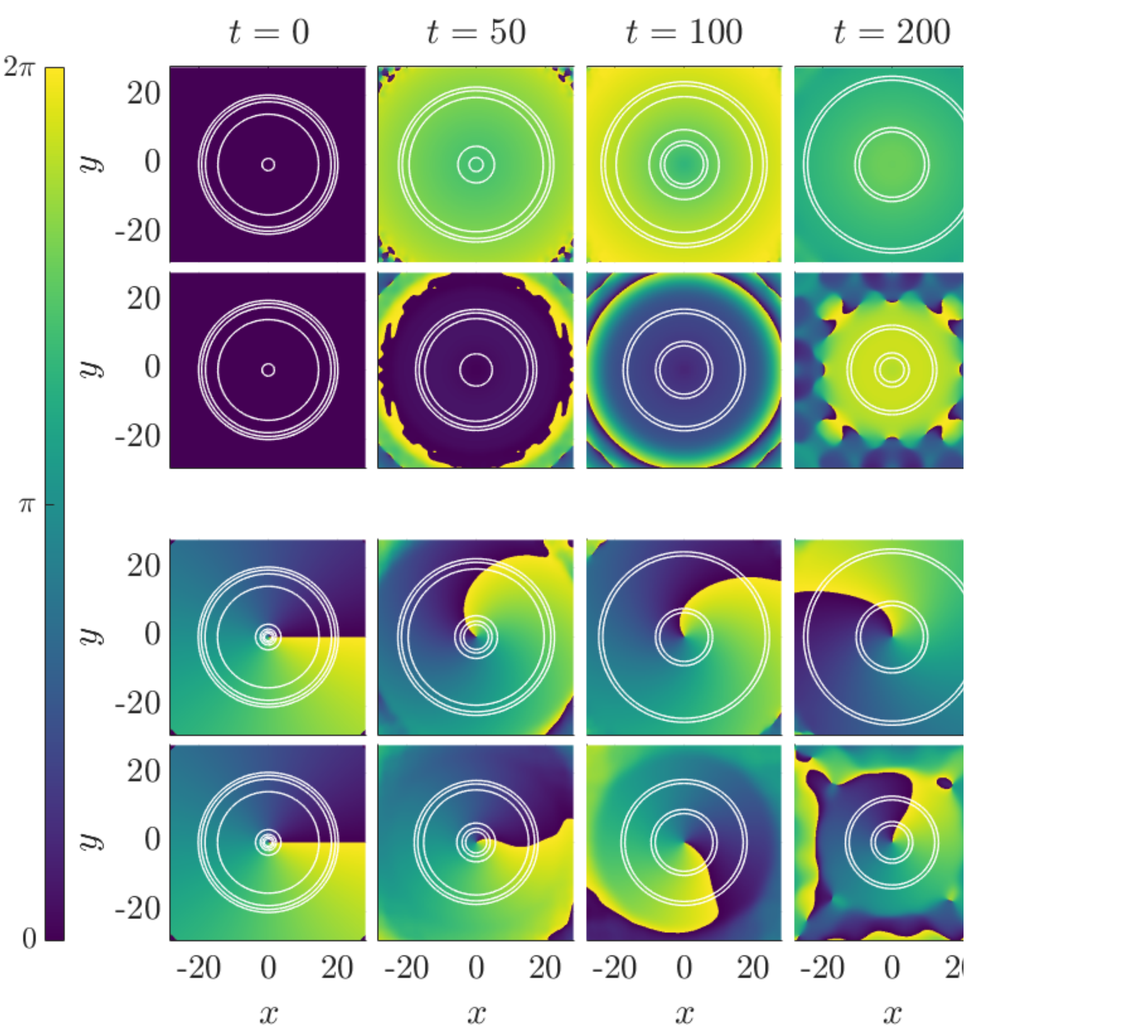}
  \caption{Phases and superimposed density contours of the time evolution of ground states with zero or one vortex in a harmonic trap with a 
  Gaussian ($V_0 = 0.2$) for $N = 1000$ and $\omega = 0.04$. Rows one and three show evolutions without three-body losses $L_3 = 0$, 
  and rows two and four with
$L_3 = 0.01$. The harmonic trap frequency is linearly decreased in time
such that it is zero at $t = 100$ while leaving the Gaussian unchanged.}
  \label{fig:panelsexp}
\end{figure}
Intriguingly, in this ideal case where conservation laws are intact, the
droplets stay stable even after the trap is fully turned off while still
carrying angular momentum in the form of vortices and rigid-body rotation. Additionally, the shape of the droplets is deformed according to the number of vortices they contain.
For a condensate in a trap with the Gaussian discussed previously, we
consider a similar release, but now leaving the Gaussian even after
the harmonic trap is turned off. Density contours and phases for the release of zero- and unit-vortex states in such a setup are displayed in
Fig.~\ref{fig:panelsexp}, where we have also included a comparison with the
corresponding systems including three-body losses with $L_3 =
0.01$. For both the cases with and without a vortex, the droplet is pinned to the remaining Gaussian, and they stay metastable and self-bound even when the symmetry-breaking three-body losses term is present.

In order to relate our results to experimental values, let us consider
$^{39}$K atoms tightly confined by a harmonic trap in the transversal
direction with an oscillator length $\ell_z = 0.1\mu\mathrm{m}$, and three
dimensional s-wave scattering lengths equal to $a_{\uparrow\downarrow} = -50.0a_0$ and $a = 50.5a_0$ for the inter- and intraspecies
interactions, respectively ($a_0$ is the Bohr radius). Note that these
values for the scattering lengths correspond to the stable gas phase in
three dimensions~\cite{cabrera2018}; the liquefaction is due to
the transition to two dimensions~\cite{petrov2016}. This choice, when transformed from the dimensionless parameters used above, corresponds roughly to $\omega\sim 10
\mathrm{Hz}$, $N \sim 10^5$ and $n\sim 10^{14} \mathrm{cm}^{-3}$, with
units of time and space in $\sim \mathrm{ms}$ and $\sim\mu\mathrm{m}$,
respectively~\cite{petrov2001}. The value used for the three-body
losses in Fig.~\ref{fig:panelsexp} corresponds approximately to
$10^{-27}\mathrm{cm}^6/\mathrm{s}$.

In conclusion, binary self-bound bosonic droplets as realized in recent experiments with potassium~\cite{semeghini2018,cabrera2018} 
show the formation of vortices in a way similar to scalar BECs with weak short-range interactions, but with the additon of a deformation to the droplets' shape. In order to stabilize these droplets and prevent their decay into fragments (such as they were found both in two and three dimensions, see ~\cite{li2018,kartashov2018}), we found that it is crucial to first stabilize the droplets by a weak harmonic confinement, chosen such that it barely confines the self-bound droplet. When switching off the trap rotation and slowly releasing the droplet, cusps in the yrast line lead to rotational ground states that can generate rotating droplets containing multiple singly-quantized vortices. To study a more pertubation-resistant system, we considered a binary BEC trapped in a mexican-hat potential, where we found the existence of a metastable persistent current that potentially could be utilized in order to produce droplets carrying angular momentum in the form of vortices. The findings presented here should be in the range of present experiments
for binary condensates, and we expect that similarily metastable persistent currents may occur in dipolar quantum droplets. While the present analysis made use of the two-dimensional 
extended Gross-Pitaevskii approach~\cite{petrov2016},
it will also be interesting to study the crossover between two and three dimensions 
~\cite{ilg2018,kartashov2018} from the perspective of vortex formation.

\bigskip

\begin{acknowledgments}
{\it Acknowledgements.} We thank in particular G. Kavoulakis for his help and useful comments at the initial stage of the project. We also thank
J. Bengtsson, J. Bjerlin, G. Eriksson, B. Mottelson and  R. Sachdeva for discussions.  This work is financially supported by The Swedish Research Council and the Knut and Alice Wallenberg Foundation. 
\end{acknowledgments}

\bibliography{droplets.bib}

\begin{thebibliography}{54}%
\makeatletter
\providecommand \@ifxundefined [1]{%
 \@ifx{#1\undefined}
}%
\providecommand \@ifnum [1]{%
 \ifnum #1\expandafter \@firstoftwo
 \else \expandafter \@secondoftwo
 \fi
}%
\providecommand \@ifx [1]{%
 \ifx #1\expandafter \@firstoftwo
 \else \expandafter \@secondoftwo
 \fi
}%
\providecommand \natexlab [1]{#1}%
\providecommand \enquote  [1]{``#1''}%
\providecommand \bibnamefont  [1]{#1}%
\providecommand \bibfnamefont [1]{#1}%
\providecommand \citenamefont [1]{#1}%
\providecommand \href@noop [0]{\@secondoftwo}%
\providecommand \href [0]{\begingroup \@sanitize@url \@href}%
\providecommand \@href[1]{\@@startlink{#1}\@@href}%
\providecommand \@@href[1]{\endgroup#1\@@endlink}%
\providecommand \@sanitize@url [0]{\catcode `\\12\catcode `\$12\catcode
  `\&12\catcode `\#12\catcode `\^12\catcode `\_12\catcode `\%12\relax}%
\providecommand \@@startlink[1]{}%
\providecommand \@@endlink[0]{}%
\providecommand \url  [0]{\begingroup\@sanitize@url \@url }%
\providecommand \@url [1]{\endgroup\@href {#1}{\urlprefix }}%
\providecommand \urlprefix  [0]{URL }%
\providecommand \Eprint [0]{\href }%
\providecommand \doibase [0]{http://dx.doi.org/}%
\providecommand \selectlanguage [0]{\@gobble}%
\providecommand \bibinfo  [0]{\@secondoftwo}%
\providecommand \bibfield  [0]{\@secondoftwo}%
\providecommand \translation [1]{[#1]}%
\providecommand \BibitemOpen [0]{}%
\providecommand \bibitemStop [0]{}%
\providecommand \bibitemNoStop [0]{.\EOS\space}%
\providecommand \EOS [0]{\spacefactor3000\relax}%
\providecommand \BibitemShut  [1]{\csname bibitem#1\endcsname}%
\let\auto@bib@innerbib\@empty
\bibitem [{\citenamefont {Bohr}\ and\ \citenamefont
  {Mottelson}(1998)}]{BohrMottelson}%
  \BibitemOpen
  \bibfield  {author} {\bibinfo {author} {\bibfnamefont {A.}~\bibnamefont
  {Bohr}}\ and\ \bibinfo {author} {\bibfnamefont {B.}~\bibnamefont
  {Mottelson}},\ }\href@noop {} {\emph {\bibinfo {title} {Nuclear
  Structure}}},\ \bibinfo {series} {Nuclear Structure}\ No.\ \bibinfo {number}
  {v. 1}\ (\bibinfo  {publisher} {World Scientific},\ \bibinfo {year}
  {1998})\BibitemShut {NoStop}%
\bibitem [{\citenamefont {Donelly}(1991)}]{donelly1991}%
  \BibitemOpen
  \bibfield  {author} {\bibinfo {author} {\bibfnamefont {R.}~\bibnamefont
  {Donelly}},\ }\href@noop {} {\emph {\bibinfo {title} {Quantized vortices in
  Helium II}}},\ \bibinfo {number} {v. 3}\ (\bibinfo  {publisher} {Cambridge
  University Press},\ \bibinfo {year} {1991})\BibitemShut {NoStop}%
\bibitem [{\citenamefont {Toennies}\ and\ \citenamefont
  {Vilesov}(2004)}]{toennies2004}%
  \BibitemOpen
  \bibfield  {author} {\bibinfo {author} {\bibfnamefont {J.}~\bibnamefont
  {Toennies}}\ and\ \bibinfo {author} {\bibfnamefont {A.}~\bibnamefont
  {Vilesov}},\ }\href {\doibase 10.1002/anie.200300611} {\bibfield  {journal}
  {\bibinfo  {journal} {Angew. Chem. Phys.}\ }\textbf {\bibinfo {volume}
  {43}},\ \bibinfo {pages} {2622} (\bibinfo {year} {2004})}\BibitemShut
  {NoStop}%
\bibitem [{\citenamefont {Ancilotto}\ \emph
  {et~al.}(2018{\natexlab{a}})\citenamefont {Ancilotto}, \citenamefont
  {Barranco},\ and\ \citenamefont {Pi}}]{ancilotto2018}%
  \BibitemOpen
  \bibfield  {author} {\bibinfo {author} {\bibfnamefont {F.}~\bibnamefont
  {Ancilotto}}, \bibinfo {author} {\bibfnamefont {M.}~\bibnamefont {Barranco}},
  \ and\ \bibinfo {author} {\bibfnamefont {M.}~\bibnamefont {Pi}},\ }\href
  {\doibase 10.1103/PhysRevB.97.184515} {\bibfield  {journal} {\bibinfo
  {journal} {Phys. Rev. B}\ }\textbf {\bibinfo {volume} {97}},\ \bibinfo
  {pages} {184515} (\bibinfo {year} {2018}{\natexlab{a}})}\BibitemShut
  {NoStop}%
\bibitem [{\citenamefont {Kadau}\ \emph {et~al.}(2016)\citenamefont {Kadau},
  \citenamefont {Schmitt}, \citenamefont {Wenzel}, \citenamefont {Wink},
  \citenamefont {Maier}, \citenamefont {Ferrier-Barbut},\ and\ \citenamefont
  {Pfau}}]{kadau2016}%
  \BibitemOpen
  \bibfield  {author} {\bibinfo {author} {\bibfnamefont {H.}~\bibnamefont
  {Kadau}}, \bibinfo {author} {\bibfnamefont {M.}~\bibnamefont {Schmitt}},
  \bibinfo {author} {\bibfnamefont {M.}~\bibnamefont {Wenzel}}, \bibinfo
  {author} {\bibfnamefont {C.}~\bibnamefont {Wink}}, \bibinfo {author}
  {\bibfnamefont {T.}~\bibnamefont {Maier}}, \bibinfo {author} {\bibfnamefont
  {I.}~\bibnamefont {Ferrier-Barbut}}, \ and\ \bibinfo {author} {\bibfnamefont
  {T.}~\bibnamefont {Pfau}},\ }\href {https://doi.org/10.1038/nature16485}
  {\bibfield  {journal} {\bibinfo  {journal} {Nature}\ }\textbf {\bibinfo
  {volume} {530}},\ \bibinfo {pages} {194} (\bibinfo {year}
  {2016})}\BibitemShut {NoStop}%
\bibitem [{\citenamefont {Schmitt}\ \emph {et~al.}(2016)\citenamefont
  {Schmitt}, \citenamefont {Wenzel}, \citenamefont {B{\"o}ttcher},
  \citenamefont {Ferrier-Barbut},\ and\ \citenamefont {Pfau}}]{schmitt2016}%
  \BibitemOpen
  \bibfield  {author} {\bibinfo {author} {\bibfnamefont {M.}~\bibnamefont
  {Schmitt}}, \bibinfo {author} {\bibfnamefont {M.}~\bibnamefont {Wenzel}},
  \bibinfo {author} {\bibfnamefont {F.}~\bibnamefont {B{\"o}ttcher}}, \bibinfo
  {author} {\bibfnamefont {I.}~\bibnamefont {Ferrier-Barbut}}, \ and\ \bibinfo
  {author} {\bibfnamefont {T.}~\bibnamefont {Pfau}},\ }\href
  {https://doi.org/10.1038/nature20126} {\bibfield  {journal} {\bibinfo
  {journal} {Nature}\ }\textbf {\bibinfo {volume} {539}},\ \bibinfo {pages}
  {259} (\bibinfo {year} {2016})}\BibitemShut {NoStop}%
\bibitem [{\citenamefont {Ferrier-Barbut}\ \emph
  {et~al.}(2016{\natexlab{a}})\citenamefont {Ferrier-Barbut}, \citenamefont
  {Kadau}, \citenamefont {Schmitt}, \citenamefont {Wenzel},\ and\ \citenamefont
  {Pfau}}]{ferrierbarbut2016}%
  \BibitemOpen
  \bibfield  {author} {\bibinfo {author} {\bibfnamefont {I.}~\bibnamefont
  {Ferrier-Barbut}}, \bibinfo {author} {\bibfnamefont {H.}~\bibnamefont
  {Kadau}}, \bibinfo {author} {\bibfnamefont {M.}~\bibnamefont {Schmitt}},
  \bibinfo {author} {\bibfnamefont {M.}~\bibnamefont {Wenzel}}, \ and\ \bibinfo
  {author} {\bibfnamefont {T.}~\bibnamefont {Pfau}},\ }\href {\doibase
  10.1103/PhysRevLett.116.215301} {\bibfield  {journal} {\bibinfo  {journal}
  {Phys. Rev. Lett.}\ }\textbf {\bibinfo {volume} {116}},\ \bibinfo {pages}
  {215301} (\bibinfo {year} {2016}{\natexlab{a}})}\BibitemShut {NoStop}%
\bibitem [{\citenamefont {Ferrier-Barbut}\ \emph
  {et~al.}(2016{\natexlab{b}})\citenamefont {Ferrier-Barbut}, \citenamefont
  {Schmitt}, \citenamefont {Wenzel}, \citenamefont {Kadau},\ and\ \citenamefont
  {Pfau}}]{ferrierbarbut2016b}%
  \BibitemOpen
  \bibfield  {author} {\bibinfo {author} {\bibfnamefont {I.}~\bibnamefont
  {Ferrier-Barbut}}, \bibinfo {author} {\bibfnamefont {M.}~\bibnamefont
  {Schmitt}}, \bibinfo {author} {\bibfnamefont {M.}~\bibnamefont {Wenzel}},
  \bibinfo {author} {\bibfnamefont {H.}~\bibnamefont {Kadau}}, \ and\ \bibinfo
  {author} {\bibfnamefont {T.}~\bibnamefont {Pfau}},\ }\href
  {https://iopscience.iop.org/article/10.1088/0953-4075/49/21/214004/meta}
  {\bibfield  {journal} {\bibinfo  {journal} {J. Phys. B}\ }\textbf {\bibinfo
  {volume} {49}},\ \bibinfo {pages} {214004} (\bibinfo {year}
  {2016}{\natexlab{b}})}\BibitemShut {NoStop}%
\bibitem [{\citenamefont {Ferrier-Barbut}\ \emph {et~al.}(2018)\citenamefont
  {Ferrier-Barbut}, \citenamefont {Wenzel}, \citenamefont {B\"ottcher},
  \citenamefont {Langen}, \citenamefont {Isoard}, \citenamefont {Stringari},\
  and\ \citenamefont {Pfau}}]{ferrierbarbut2018}%
  \BibitemOpen
  \bibfield  {author} {\bibinfo {author} {\bibfnamefont {I.}~\bibnamefont
  {Ferrier-Barbut}}, \bibinfo {author} {\bibfnamefont {M.}~\bibnamefont
  {Wenzel}}, \bibinfo {author} {\bibfnamefont {F.}~\bibnamefont {B\"ottcher}},
  \bibinfo {author} {\bibfnamefont {T.}~\bibnamefont {Langen}}, \bibinfo
  {author} {\bibfnamefont {M.}~\bibnamefont {Isoard}}, \bibinfo {author}
  {\bibfnamefont {S.}~\bibnamefont {Stringari}}, \ and\ \bibinfo {author}
  {\bibfnamefont {T.}~\bibnamefont {Pfau}},\ }\href {\doibase
  10.1103/PhysRevLett.120.160402} {\bibfield  {journal} {\bibinfo  {journal}
  {Phys. Rev. Lett.}\ }\textbf {\bibinfo {volume} {120}},\ \bibinfo {pages}
  {160402} (\bibinfo {year} {2018})}\BibitemShut {NoStop}%
\bibitem [{\citenamefont {Chomaz}\ \emph {et~al.}(2016)\citenamefont {Chomaz},
  \citenamefont {Baier}, \citenamefont {Petter}, \citenamefont {Mark},
  \citenamefont {W\"achtler}, \citenamefont {Santos},\ and\ \citenamefont
  {Ferlaino}}]{chomaz2016}%
  \BibitemOpen
  \bibfield  {author} {\bibinfo {author} {\bibfnamefont {L.}~\bibnamefont
  {Chomaz}}, \bibinfo {author} {\bibfnamefont {S.}~\bibnamefont {Baier}},
  \bibinfo {author} {\bibfnamefont {D.}~\bibnamefont {Petter}}, \bibinfo
  {author} {\bibfnamefont {M.~J.}\ \bibnamefont {Mark}}, \bibinfo {author}
  {\bibfnamefont {F.}~\bibnamefont {W\"achtler}}, \bibinfo {author}
  {\bibfnamefont {L.}~\bibnamefont {Santos}}, \ and\ \bibinfo {author}
  {\bibfnamefont {F.}~\bibnamefont {Ferlaino}},\ }\href {\doibase
  10.1103/PhysRevX.6.041039} {\bibfield  {journal} {\bibinfo  {journal} {Phys.
  Rev. X}\ }\textbf {\bibinfo {volume} {6}},\ \bibinfo {pages} {041039}
  (\bibinfo {year} {2016})}\BibitemShut {NoStop}%
\bibitem [{\citenamefont {Lu}\ \emph {et~al.}(2010)\citenamefont {Lu},
  \citenamefont {Youn},\ and\ \citenamefont {Lev}}]{lu2010}%
  \BibitemOpen
  \bibfield  {author} {\bibinfo {author} {\bibfnamefont {M.}~\bibnamefont
  {Lu}}, \bibinfo {author} {\bibfnamefont {S.~H.}\ \bibnamefont {Youn}}, \ and\
  \bibinfo {author} {\bibfnamefont {B.~L.}\ \bibnamefont {Lev}},\ }\href
  {\doibase 10.1103/PhysRevLett.104.063001} {\bibfield  {journal} {\bibinfo
  {journal} {Phys. Rev. Lett.}\ }\textbf {\bibinfo {volume} {104}},\ \bibinfo
  {pages} {063001} (\bibinfo {year} {2010})}\BibitemShut {NoStop}%
\bibitem [{\citenamefont {Lu}\ \emph {et~al.}(2011)\citenamefont {Lu},
  \citenamefont {Burdick}, \citenamefont {Youn},\ and\ \citenamefont
  {Lev}}]{lu2011}%
  \BibitemOpen
  \bibfield  {author} {\bibinfo {author} {\bibfnamefont {M.}~\bibnamefont
  {Lu}}, \bibinfo {author} {\bibfnamefont {N.~Q.}\ \bibnamefont {Burdick}},
  \bibinfo {author} {\bibfnamefont {S.~H.}\ \bibnamefont {Youn}}, \ and\
  \bibinfo {author} {\bibfnamefont {B.~L.}\ \bibnamefont {Lev}},\ }\href
  {\doibase 10.1103/PhysRevLett.107.190401} {\bibfield  {journal} {\bibinfo
  {journal} {Phys. Rev. Lett.}\ }\textbf {\bibinfo {volume} {107}},\ \bibinfo
  {pages} {190401} (\bibinfo {year} {2011})}\BibitemShut {NoStop}%
\bibitem [{\citenamefont {Aikawa}\ \emph {et~al.}(2012)\citenamefont {Aikawa},
  \citenamefont {Frisch}, \citenamefont {Mark}, \citenamefont {Baier},
  \citenamefont {Rietzler}, \citenamefont {Grimm},\ and\ \citenamefont
  {Ferlaino}}]{aikawa2012}%
  \BibitemOpen
  \bibfield  {author} {\bibinfo {author} {\bibfnamefont {K.}~\bibnamefont
  {Aikawa}}, \bibinfo {author} {\bibfnamefont {A.}~\bibnamefont {Frisch}},
  \bibinfo {author} {\bibfnamefont {M.}~\bibnamefont {Mark}}, \bibinfo {author}
  {\bibfnamefont {S.}~\bibnamefont {Baier}}, \bibinfo {author} {\bibfnamefont
  {A.}~\bibnamefont {Rietzler}}, \bibinfo {author} {\bibfnamefont
  {R.}~\bibnamefont {Grimm}}, \ and\ \bibinfo {author} {\bibfnamefont
  {F.}~\bibnamefont {Ferlaino}},\ }\href {\doibase
  10.1103/PhysRevLett.108.210401} {\bibfield  {journal} {\bibinfo  {journal}
  {Phys. Rev. Lett.}\ }\textbf {\bibinfo {volume} {108}},\ \bibinfo {pages}
  {210401} (\bibinfo {year} {2012})}\BibitemShut {NoStop}%
\bibitem [{\citenamefont {Cabrera}\ \emph {et~al.}(2018)\citenamefont
  {Cabrera}, \citenamefont {Tanzi}, \citenamefont {Sanz}, \citenamefont
  {Naylor}, \citenamefont {Thomas}, \citenamefont {Cheiney},\ and\
  \citenamefont {Tarruell}}]{cabrera2018}%
  \BibitemOpen
  \bibfield  {author} {\bibinfo {author} {\bibfnamefont {C.}~\bibnamefont
  {Cabrera}}, \bibinfo {author} {\bibfnamefont {L.}~\bibnamefont {Tanzi}},
  \bibinfo {author} {\bibfnamefont {J.}~\bibnamefont {Sanz}}, \bibinfo {author}
  {\bibfnamefont {B.}~\bibnamefont {Naylor}}, \bibinfo {author} {\bibfnamefont
  {P.}~\bibnamefont {Thomas}}, \bibinfo {author} {\bibfnamefont
  {P.}~\bibnamefont {Cheiney}}, \ and\ \bibinfo {author} {\bibfnamefont
  {L.}~\bibnamefont {Tarruell}},\ }\href {\doibase 10.1126/science.aao5686}
  {\bibfield  {journal} {\bibinfo  {journal} {Science}\ }\textbf {\bibinfo
  {volume} {359}},\ \bibinfo {pages} {301} (\bibinfo {year}
  {2018})}\BibitemShut {NoStop}%
\bibitem [{\citenamefont {Semeghini}\ \emph {et~al.}(2018)\citenamefont
  {Semeghini}, \citenamefont {Ferioli}, \citenamefont {Masi}, \citenamefont
  {Mazzinghi}, \citenamefont {Wolswijk}, \citenamefont {Minardi}, \citenamefont
  {Modugno}, \citenamefont {Modugno}, \citenamefont {Inguscio},\ and\
  \citenamefont {Fattori}}]{semeghini2018}%
  \BibitemOpen
  \bibfield  {author} {\bibinfo {author} {\bibfnamefont {G.}~\bibnamefont
  {Semeghini}}, \bibinfo {author} {\bibfnamefont {G.}~\bibnamefont {Ferioli}},
  \bibinfo {author} {\bibfnamefont {L.}~\bibnamefont {Masi}}, \bibinfo {author}
  {\bibfnamefont {C.}~\bibnamefont {Mazzinghi}}, \bibinfo {author}
  {\bibfnamefont {L.}~\bibnamefont {Wolswijk}}, \bibinfo {author}
  {\bibfnamefont {F.}~\bibnamefont {Minardi}}, \bibinfo {author} {\bibfnamefont
  {M.}~\bibnamefont {Modugno}}, \bibinfo {author} {\bibfnamefont
  {G.}~\bibnamefont {Modugno}}, \bibinfo {author} {\bibfnamefont
  {M.}~\bibnamefont {Inguscio}}, \ and\ \bibinfo {author} {\bibfnamefont
  {M.}~\bibnamefont {Fattori}},\ }\href {\doibase
  10.1103/PhysRevLett.120.235301} {\bibfield  {journal} {\bibinfo  {journal}
  {Phys. Rev. Lett.}\ }\textbf {\bibinfo {volume} {120}},\ \bibinfo {pages}
  {235301} (\bibinfo {year} {2018})}\BibitemShut {NoStop}%
\bibitem [{\citenamefont {Petrov}(2015)}]{petrov2015}%
  \BibitemOpen
  \bibfield  {author} {\bibinfo {author} {\bibfnamefont {D.~S.}\ \bibnamefont
  {Petrov}},\ }\href {\doibase 10.1103/PhysRevLett.115.155302} {\bibfield
  {journal} {\bibinfo  {journal} {Phys. Rev. Lett.}\ }\textbf {\bibinfo
  {volume} {115}},\ \bibinfo {pages} {155302} (\bibinfo {year}
  {2015})}\BibitemShut {NoStop}%
\bibitem [{\citenamefont {Petrov}\ and\ \citenamefont
  {Astrakharchik}(2016)}]{petrov2016}%
  \BibitemOpen
  \bibfield  {author} {\bibinfo {author} {\bibfnamefont {D.~S.}\ \bibnamefont
  {Petrov}}\ and\ \bibinfo {author} {\bibfnamefont {G.~E.}\ \bibnamefont
  {Astrakharchik}},\ }\href {\doibase 10.1103/PhysRevLett.117.100401}
  {\bibfield  {journal} {\bibinfo  {journal} {Phys. Rev. Lett.}\ }\textbf
  {\bibinfo {volume} {117}},\ \bibinfo {pages} {100401} (\bibinfo {year}
  {2016})}\BibitemShut {NoStop}%
\bibitem [{\citenamefont {Bulgac}(2002)}]{bulgac2002}%
  \BibitemOpen
  \bibfield  {author} {\bibinfo {author} {\bibfnamefont {A.}~\bibnamefont
  {Bulgac}},\ }\href {\doibase 10.1103/PhysRevLett.89.050402} {\bibfield
  {journal} {\bibinfo  {journal} {Phys. Rev. Lett.}\ }\textbf {\bibinfo
  {volume} {89}},\ \bibinfo {pages} {050402} (\bibinfo {year}
  {2002})}\BibitemShut {NoStop}%
\bibitem [{\citenamefont {Hammer}\ and\ \citenamefont
  {Son}(2004)}]{hammer2004}%
  \BibitemOpen
  \bibfield  {author} {\bibinfo {author} {\bibfnamefont {H.-W.}\ \bibnamefont
  {Hammer}}\ and\ \bibinfo {author} {\bibfnamefont {D.~T.}\ \bibnamefont
  {Son}},\ }\href {\doibase 10.1103/PhysRevLett.93.250408} {\bibfield
  {journal} {\bibinfo  {journal} {Phys. Rev. Lett.}\ }\textbf {\bibinfo
  {volume} {93}},\ \bibinfo {pages} {250408} (\bibinfo {year}
  {2004})}\BibitemShut {NoStop}%
\bibitem [{\citenamefont {Lee}\ \emph {et~al.}(1957)\citenamefont {Lee},
  \citenamefont {Huang},\ and\ \citenamefont {Yang}}]{lhy1957}%
  \BibitemOpen
  \bibfield  {author} {\bibinfo {author} {\bibfnamefont {T.~D.}\ \bibnamefont
  {Lee}}, \bibinfo {author} {\bibfnamefont {K.}~\bibnamefont {Huang}}, \ and\
  \bibinfo {author} {\bibfnamefont {C.~N.}\ \bibnamefont {Yang}},\ }\href
  {\doibase 10.1103/PhysRev.106.1135} {\bibfield  {journal} {\bibinfo
  {journal} {Phys. Rev.}\ }\textbf {\bibinfo {volume} {106}},\ \bibinfo {pages}
  {1135} (\bibinfo {year} {1957})}\BibitemShut {NoStop}%
\bibitem [{\citenamefont {Knight}\ \emph {et~al.}(1984)\citenamefont {Knight},
  \citenamefont {Clemenger}, \citenamefont {de~Heer}, \citenamefont {Saunders},
  \citenamefont {Chou},\ and\ \citenamefont {Cohen}}]{knight1984}%
  \BibitemOpen
  \bibfield  {author} {\bibinfo {author} {\bibfnamefont {W.~D.}\ \bibnamefont
  {Knight}}, \bibinfo {author} {\bibfnamefont {K.}~\bibnamefont {Clemenger}},
  \bibinfo {author} {\bibfnamefont {W.~A.}\ \bibnamefont {de~Heer}}, \bibinfo
  {author} {\bibfnamefont {W.~A.}\ \bibnamefont {Saunders}}, \bibinfo {author}
  {\bibfnamefont {M.~Y.}\ \bibnamefont {Chou}}, \ and\ \bibinfo {author}
  {\bibfnamefont {M.~L.}\ \bibnamefont {Cohen}},\ }\href {\doibase
  10.1103/PhysRevLett.52.2141} {\bibfield  {journal} {\bibinfo  {journal}
  {Phys. Rev. Lett.}\ }\textbf {\bibinfo {volume} {52}},\ \bibinfo {pages}
  {2141} (\bibinfo {year} {1984})}\BibitemShut {NoStop}%
\bibitem [{\citenamefont {Nishioka}\ \emph {et~al.}(1990)\citenamefont
  {Nishioka}, \citenamefont {Hansen},\ and\ \citenamefont
  {Mottelson}}]{nishioka1990}%
  \BibitemOpen
  \bibfield  {author} {\bibinfo {author} {\bibfnamefont {H.}~\bibnamefont
  {Nishioka}}, \bibinfo {author} {\bibfnamefont {K.}~\bibnamefont {Hansen}}, \
  and\ \bibinfo {author} {\bibfnamefont {B.~R.}\ \bibnamefont {Mottelson}},\
  }\href {\doibase 10.1103/PhysRevB.42.9377} {\bibfield  {journal} {\bibinfo
  {journal} {Phys. Rev. B}\ }\textbf {\bibinfo {volume} {42}},\ \bibinfo
  {pages} {9377} (\bibinfo {year} {1990})}\BibitemShut {NoStop}%
\bibitem [{\citenamefont {Koskinen}\ \emph {et~al.}(1995)\citenamefont
  {Koskinen}, \citenamefont {Lipas},\ and\ \citenamefont
  {Manninen}}]{koskinen1995}%
  \BibitemOpen
  \bibfield  {author} {\bibinfo {author} {\bibfnamefont {M.}~\bibnamefont
  {Koskinen}}, \bibinfo {author} {\bibfnamefont {P.~O.}\ \bibnamefont {Lipas}},
  \ and\ \bibinfo {author} {\bibfnamefont {M.}~\bibnamefont {Manninen}},\
  }\href {\doibase 10.1007/BF01745532} {\bibfield  {journal} {\bibinfo
  {journal} {Zeitschrift f{\"u}r Physik D Atoms, Molecules and Clusters}\
  }\textbf {\bibinfo {volume} {35}},\ \bibinfo {pages} {285} (\bibinfo {year}
  {1995})}\BibitemShut {NoStop}%
\bibitem [{\citenamefont {Baillie}\ \emph {et~al.}(2016)\citenamefont
  {Baillie}, \citenamefont {Wilson}, \citenamefont {Bisset},\ and\
  \citenamefont {Blakie}}]{baillie2016}%
  \BibitemOpen
  \bibfield  {author} {\bibinfo {author} {\bibfnamefont {D.}~\bibnamefont
  {Baillie}}, \bibinfo {author} {\bibfnamefont {R.~M.}\ \bibnamefont {Wilson}},
  \bibinfo {author} {\bibfnamefont {R.~N.}\ \bibnamefont {Bisset}}, \ and\
  \bibinfo {author} {\bibfnamefont {P.~B.}\ \bibnamefont {Blakie}},\ }\href
  {\doibase 10.1103/PhysRevA.94.021602} {\bibfield  {journal} {\bibinfo
  {journal} {Phys. Rev. A}\ }\textbf {\bibinfo {volume} {94}},\ \bibinfo
  {pages} {021602} (\bibinfo {year} {2016})}\BibitemShut {NoStop}%
\bibitem [{\citenamefont {W\"achtler}\ and\ \citenamefont
  {Santos}(2016{\natexlab{a}})}]{wachtler2016}%
  \BibitemOpen
  \bibfield  {author} {\bibinfo {author} {\bibfnamefont {F.}~\bibnamefont
  {W\"achtler}}\ and\ \bibinfo {author} {\bibfnamefont {L.}~\bibnamefont
  {Santos}},\ }\href {\doibase 10.1103/PhysRevA.93.061603} {\bibfield
  {journal} {\bibinfo  {journal} {Phys. Rev. A}\ }\textbf {\bibinfo {volume}
  {93}},\ \bibinfo {pages} {061603} (\bibinfo {year}
  {2016}{\natexlab{a}})}\BibitemShut {NoStop}%
\bibitem [{\citenamefont {W\"achtler}\ and\ \citenamefont
  {Santos}(2016{\natexlab{b}})}]{wachtler2016b}%
  \BibitemOpen
  \bibfield  {author} {\bibinfo {author} {\bibfnamefont {F.}~\bibnamefont
  {W\"achtler}}\ and\ \bibinfo {author} {\bibfnamefont {L.}~\bibnamefont
  {Santos}},\ }\href {\doibase 10.1103/PhysRevA.94.043618} {\bibfield
  {journal} {\bibinfo  {journal} {Phys. Rev. A}\ }\textbf {\bibinfo {volume}
  {94}},\ \bibinfo {pages} {043618} (\bibinfo {year}
  {2016}{\natexlab{b}})}\BibitemShut {NoStop}%
\bibitem [{\citenamefont {Bisset}\ \emph {et~al.}(2016)\citenamefont {Bisset},
  \citenamefont {Wilson}, \citenamefont {Baillie},\ and\ \citenamefont
  {Blakie}}]{bisset2016}%
  \BibitemOpen
  \bibfield  {author} {\bibinfo {author} {\bibfnamefont {R.~N.}\ \bibnamefont
  {Bisset}}, \bibinfo {author} {\bibfnamefont {R.~M.}\ \bibnamefont {Wilson}},
  \bibinfo {author} {\bibfnamefont {D.}~\bibnamefont {Baillie}}, \ and\
  \bibinfo {author} {\bibfnamefont {P.~B.}\ \bibnamefont {Blakie}},\ }\href
  {\doibase 10.1103/PhysRevA.94.033619} {\bibfield  {journal} {\bibinfo
  {journal} {Phys. Rev. A}\ }\textbf {\bibinfo {volume} {94}},\ \bibinfo
  {pages} {033619} (\bibinfo {year} {2016})}\BibitemShut {NoStop}%
\bibitem [{\citenamefont {Baillie}\ \emph {et~al.}(2017)\citenamefont
  {Baillie}, \citenamefont {Wilson},\ and\ \citenamefont
  {Blakie}}]{baillie2017}%
  \BibitemOpen
  \bibfield  {author} {\bibinfo {author} {\bibfnamefont {D.}~\bibnamefont
  {Baillie}}, \bibinfo {author} {\bibfnamefont {R.~M.}\ \bibnamefont {Wilson}},
  \ and\ \bibinfo {author} {\bibfnamefont {P.~B.}\ \bibnamefont {Blakie}},\
  }\href {\doibase 10.1103/PhysRevLett.119.255302} {\bibfield  {journal}
  {\bibinfo  {journal} {Phys. Rev. Lett.}\ }\textbf {\bibinfo {volume} {119}},\
  \bibinfo {pages} {255302} (\bibinfo {year} {2017})}\BibitemShut {NoStop}%
\bibitem [{\citenamefont {Saito}(2016)}]{saito2016}%
  \BibitemOpen
  \bibfield  {author} {\bibinfo {author} {\bibfnamefont {H.}~\bibnamefont
  {Saito}},\ }\href {\doibase 10.7566/JPSJ.85.053001} {\bibfield  {journal}
  {\bibinfo  {journal} {J. Phys. Soc. Jpn.}\ }\textbf {\bibinfo {volume}
  {85}},\ \bibinfo {pages} {053001} (\bibinfo {year} {2016})}\BibitemShut
  {NoStop}%
\bibitem [{\citenamefont {Macia}\ \emph {et~al.}(2016)\citenamefont {Macia},
  \citenamefont {S\'anchez-Baena}, \citenamefont {Boronat},\ and\ \citenamefont
  {Mazzanti}}]{macia2016}%
  \BibitemOpen
  \bibfield  {author} {\bibinfo {author} {\bibfnamefont {A.}~\bibnamefont
  {Macia}}, \bibinfo {author} {\bibfnamefont {J.}~\bibnamefont
  {S\'anchez-Baena}}, \bibinfo {author} {\bibfnamefont {J.}~\bibnamefont
  {Boronat}}, \ and\ \bibinfo {author} {\bibfnamefont {F.}~\bibnamefont
  {Mazzanti}},\ }\href {\doibase 10.1103/PhysRevLett.117.205301} {\bibfield
  {journal} {\bibinfo  {journal} {Phys. Rev. Lett.}\ }\textbf {\bibinfo
  {volume} {117}},\ \bibinfo {pages} {205301} (\bibinfo {year}
  {2016})}\BibitemShut {NoStop}%
\bibitem [{\citenamefont {Cinti}\ \emph {et~al.}(2017)\citenamefont {Cinti},
  \citenamefont {Cappellaro}, \citenamefont {Salasnich},\ and\ \citenamefont
  {Macr\`{\i}}}]{cinti2017}%
  \BibitemOpen
  \bibfield  {author} {\bibinfo {author} {\bibfnamefont {F.}~\bibnamefont
  {Cinti}}, \bibinfo {author} {\bibfnamefont {A.}~\bibnamefont {Cappellaro}},
  \bibinfo {author} {\bibfnamefont {L.}~\bibnamefont {Salasnich}}, \ and\
  \bibinfo {author} {\bibfnamefont {T.}~\bibnamefont {Macr\`{\i}}},\ }\href
  {\doibase 10.1103/PhysRevLett.119.215302} {\bibfield  {journal} {\bibinfo
  {journal} {Phys. Rev. Lett.}\ }\textbf {\bibinfo {volume} {119}},\ \bibinfo
  {pages} {215302} (\bibinfo {year} {2017})}\BibitemShut {NoStop}%
\bibitem [{\citenamefont {Cikojevi\ifmmode~\acute{c}\else \'{c}\fi{}}\ \emph
  {et~al.}(2018)\citenamefont {Cikojevi\ifmmode~\acute{c}\else \'{c}\fi{}},
  \citenamefont {D\ifmmode~\check{z}\else \v{z}\fi{}elalija}, \citenamefont
  {Stipanovi\ifmmode~\acute{c}\else \'{c}\fi{}}, \citenamefont {Vranje\ifmmode
  \check{s}\else \v{s}\fi{} Marki\ifmmode~\acute{c}\else \'{c}\fi{}},\ and\
  \citenamefont {Boronat}}]{cikojevic2018}%
  \BibitemOpen
  \bibfield  {author} {\bibinfo {author} {\bibfnamefont {V.}~\bibnamefont
  {Cikojevi\ifmmode~\acute{c}\else \'{c}\fi{}}}, \bibinfo {author}
  {\bibfnamefont {K.}~\bibnamefont {D\ifmmode~\check{z}\else
  \v{z}\fi{}elalija}}, \bibinfo {author} {\bibfnamefont {P.}~\bibnamefont
  {Stipanovi\ifmmode~\acute{c}\else \'{c}\fi{}}}, \bibinfo {author}
  {\bibfnamefont {L.}~\bibnamefont {Vranje\ifmmode \check{s}\else \v{s}\fi{}
  Marki\ifmmode~\acute{c}\else \'{c}\fi{}}}, \ and\ \bibinfo {author}
  {\bibfnamefont {J.}~\bibnamefont {Boronat}},\ }\href {\doibase
  10.1103/PhysRevB.97.140502} {\bibfield  {journal} {\bibinfo  {journal} {Phys.
  Rev. B}\ }\textbf {\bibinfo {volume} {97}},\ \bibinfo {pages} {140502}
  (\bibinfo {year} {2018})}\BibitemShut {NoStop}%
\bibitem [{\citenamefont {Butts}\ and\ \citenamefont
  {Rokhsar}(1999)}]{butts1999}%
  \BibitemOpen
  \bibfield  {author} {\bibinfo {author} {\bibfnamefont {D.~A.}\ \bibnamefont
  {Butts}}\ and\ \bibinfo {author} {\bibfnamefont {D.~S.}\ \bibnamefont
  {Rokhsar}},\ }\href {https://doi.org/10.1038/16865} {\bibfield  {journal}
  {\bibinfo  {journal} {Nature}\ }\textbf {\bibinfo {volume} {397}},\ \bibinfo
  {pages} {327 EP } (\bibinfo {year} {1999})}\BibitemShut {NoStop}%
\bibitem [{\citenamefont {Matthews}\ \emph {et~al.}(1999)\citenamefont
  {Matthews}, \citenamefont {Anderson}, \citenamefont {Haljan}, \citenamefont
  {Hall}, \citenamefont {Wieman},\ and\ \citenamefont
  {Cornell}}]{matthews1999}%
  \BibitemOpen
  \bibfield  {author} {\bibinfo {author} {\bibfnamefont {M.~R.}\ \bibnamefont
  {Matthews}}, \bibinfo {author} {\bibfnamefont {B.~P.}\ \bibnamefont
  {Anderson}}, \bibinfo {author} {\bibfnamefont {P.~C.}\ \bibnamefont
  {Haljan}}, \bibinfo {author} {\bibfnamefont {D.~S.}\ \bibnamefont {Hall}},
  \bibinfo {author} {\bibfnamefont {C.~E.}\ \bibnamefont {Wieman}}, \ and\
  \bibinfo {author} {\bibfnamefont {E.~A.}\ \bibnamefont {Cornell}},\ }\href
  {\doibase 10.1103/PhysRevLett.83.2498} {\bibfield  {journal} {\bibinfo
  {journal} {Phys. Rev. Lett.}\ }\textbf {\bibinfo {volume} {83}},\ \bibinfo
  {pages} {2498} (\bibinfo {year} {1999})}\BibitemShut {NoStop}%
\bibitem [{\citenamefont {Madison}\ \emph {et~al.}(2000)\citenamefont
  {Madison}, \citenamefont {Chevy}, \citenamefont {Wohlleben},\ and\
  \citenamefont {Dalibard}}]{madison2000}%
  \BibitemOpen
  \bibfield  {author} {\bibinfo {author} {\bibfnamefont {K.~W.}\ \bibnamefont
  {Madison}}, \bibinfo {author} {\bibfnamefont {F.}~\bibnamefont {Chevy}},
  \bibinfo {author} {\bibfnamefont {W.}~\bibnamefont {Wohlleben}}, \ and\
  \bibinfo {author} {\bibfnamefont {J.}~\bibnamefont {Dalibard}},\ }\href
  {\doibase 10.1103/PhysRevLett.84.806} {\bibfield  {journal} {\bibinfo
  {journal} {Phys. Rev. Lett.}\ }\textbf {\bibinfo {volume} {84}},\ \bibinfo
  {pages} {806} (\bibinfo {year} {2000})}\BibitemShut {NoStop}%
\bibitem [{\citenamefont {Chevy}\ \emph {et~al.}(2000)\citenamefont {Chevy},
  \citenamefont {Madison},\ and\ \citenamefont {Dalibard}}]{chevy2000}%
  \BibitemOpen
  \bibfield  {author} {\bibinfo {author} {\bibfnamefont {F.}~\bibnamefont
  {Chevy}}, \bibinfo {author} {\bibfnamefont {K.~W.}\ \bibnamefont {Madison}},
  \ and\ \bibinfo {author} {\bibfnamefont {J.}~\bibnamefont {Dalibard}},\
  }\href {\doibase 10.1103/PhysRevLett.85.2223} {\bibfield  {journal} {\bibinfo
   {journal} {Phys. Rev. Lett.}\ }\textbf {\bibinfo {volume} {85}},\ \bibinfo
  {pages} {2223} (\bibinfo {year} {2000})}\BibitemShut {NoStop}%
\bibitem [{\citenamefont {Kavoulakis}\ \emph {et~al.}(2000)\citenamefont
  {Kavoulakis}, \citenamefont {Mottelson},\ and\ \citenamefont
  {Pethick}}]{kavoulakis2000}%
  \BibitemOpen
  \bibfield  {author} {\bibinfo {author} {\bibfnamefont {G.~M.}\ \bibnamefont
  {Kavoulakis}}, \bibinfo {author} {\bibfnamefont {B.}~\bibnamefont
  {Mottelson}}, \ and\ \bibinfo {author} {\bibfnamefont {C.~J.}\ \bibnamefont
  {Pethick}},\ }\href {\doibase 10.1103/PhysRevA.62.063605} {\bibfield
  {journal} {\bibinfo  {journal} {Phys. Rev. A}\ }\textbf {\bibinfo {volume}
  {62}},\ \bibinfo {pages} {063605} (\bibinfo {year} {2000})}\BibitemShut
  {NoStop}%
\bibitem [{\citenamefont {Abo-Shaeer}\ \emph {et~al.}(2001)\citenamefont
  {Abo-Shaeer}, \citenamefont {Raman}, \citenamefont {Vogels},\ and\
  \citenamefont {Ketterle}}]{aboshaeer2001}%
  \BibitemOpen
  \bibfield  {author} {\bibinfo {author} {\bibfnamefont {J.~R.}\ \bibnamefont
  {Abo-Shaeer}}, \bibinfo {author} {\bibfnamefont {C.}~\bibnamefont {Raman}},
  \bibinfo {author} {\bibfnamefont {J.~M.}\ \bibnamefont {Vogels}}, \ and\
  \bibinfo {author} {\bibfnamefont {W.}~\bibnamefont {Ketterle}},\ }\href
  {\doibase 10.1126/science.1060182} {\bibfield  {journal} {\bibinfo  {journal}
  {Science}\ }\textbf {\bibinfo {volume} {292}},\ \bibinfo {pages} {476}
  (\bibinfo {year} {2001})}\BibitemShut {NoStop}%
\bibitem [{\citenamefont {Raman}\ \emph {et~al.}(2001)\citenamefont {Raman},
  \citenamefont {Abo-Shaeer}, \citenamefont {Vogels}, \citenamefont {Xu},\ and\
  \citenamefont {Ketterle}}]{raman2001}%
  \BibitemOpen
  \bibfield  {author} {\bibinfo {author} {\bibfnamefont {C.}~\bibnamefont
  {Raman}}, \bibinfo {author} {\bibfnamefont {J.~R.}\ \bibnamefont
  {Abo-Shaeer}}, \bibinfo {author} {\bibfnamefont {J.~M.}\ \bibnamefont
  {Vogels}}, \bibinfo {author} {\bibfnamefont {K.}~\bibnamefont {Xu}}, \ and\
  \bibinfo {author} {\bibfnamefont {W.}~\bibnamefont {Ketterle}},\ }\href
  {\doibase 10.1103/PhysRevLett.87.210402} {\bibfield  {journal} {\bibinfo
  {journal} {Phys. Rev. Lett.}\ }\textbf {\bibinfo {volume} {87}},\ \bibinfo
  {pages} {210402} (\bibinfo {year} {2001})}\BibitemShut {NoStop}%
\bibitem [{\citenamefont {Madison}\ \emph {et~al.}(2001)\citenamefont
  {Madison}, \citenamefont {Chevy}, \citenamefont {Bretin},\ and\ \citenamefont
  {Dalibard}}]{madison2001}%
  \BibitemOpen
  \bibfield  {author} {\bibinfo {author} {\bibfnamefont {K.~W.}\ \bibnamefont
  {Madison}}, \bibinfo {author} {\bibfnamefont {F.}~\bibnamefont {Chevy}},
  \bibinfo {author} {\bibfnamefont {V.}~\bibnamefont {Bretin}}, \ and\ \bibinfo
  {author} {\bibfnamefont {J.}~\bibnamefont {Dalibard}},\ }\href {\doibase
  10.1103/PhysRevLett.86.4443} {\bibfield  {journal} {\bibinfo  {journal}
  {Phys. Rev. Lett.}\ }\textbf {\bibinfo {volume} {86}},\ \bibinfo {pages}
  {4443} (\bibinfo {year} {2001})}\BibitemShut {NoStop}%
\bibitem [{\citenamefont {Haljan}\ \emph {et~al.}(2001)\citenamefont {Haljan},
  \citenamefont {Coddington}, \citenamefont {Engels},\ and\ \citenamefont
  {Cornell}}]{haljan2001}%
  \BibitemOpen
  \bibfield  {author} {\bibinfo {author} {\bibfnamefont {P.~C.}\ \bibnamefont
  {Haljan}}, \bibinfo {author} {\bibfnamefont {I.}~\bibnamefont {Coddington}},
  \bibinfo {author} {\bibfnamefont {P.}~\bibnamefont {Engels}}, \ and\ \bibinfo
  {author} {\bibfnamefont {E.~A.}\ \bibnamefont {Cornell}},\ }\href {\doibase
  10.1103/PhysRevLett.87.210403} {\bibfield  {journal} {\bibinfo  {journal}
  {Phys. Rev. Lett.}\ }\textbf {\bibinfo {volume} {87}},\ \bibinfo {pages}
  {210403} (\bibinfo {year} {2001})}\BibitemShut {NoStop}%
\bibitem [{\citenamefont {Fetter}(2009)}]{fetter2009}%
  \BibitemOpen
  \bibfield  {author} {\bibinfo {author} {\bibfnamefont {A.~L.}\ \bibnamefont
  {Fetter}},\ }\href {\doibase 10.1103/RevModPhys.81.647} {\bibfield  {journal}
  {\bibinfo  {journal} {Rev. Mod. Phys.}\ }\textbf {\bibinfo {volume} {81}},\
  \bibinfo {pages} {647} (\bibinfo {year} {2009})}\BibitemShut {NoStop}%
\bibitem [{\citenamefont {Saarikoski}\ \emph {et~al.}(2010)\citenamefont
  {Saarikoski}, \citenamefont {Reimann}, \citenamefont {Harju},\ and\
  \citenamefont {Manninen}}]{saarikoski2010}%
  \BibitemOpen
  \bibfield  {author} {\bibinfo {author} {\bibfnamefont {H.}~\bibnamefont
  {Saarikoski}}, \bibinfo {author} {\bibfnamefont {S.~M.}\ \bibnamefont
  {Reimann}}, \bibinfo {author} {\bibfnamefont {A.}~\bibnamefont {Harju}}, \
  and\ \bibinfo {author} {\bibfnamefont {M.}~\bibnamefont {Manninen}},\ }\href
  {\doibase 10.1103/RevModPhys.82.2785} {\bibfield  {journal} {\bibinfo
  {journal} {Rev. Mod. Phys.}\ }\textbf {\bibinfo {volume} {82}},\ \bibinfo
  {pages} {2785} (\bibinfo {year} {2010})}\BibitemShut {NoStop}%
\bibitem [{\citenamefont {Kartashov}\ \emph {et~al.}(2019)\citenamefont
  {Kartashov}, \citenamefont {Malomed},\ and\ \citenamefont
  {Torner}}]{kartashov2019}%
  \BibitemOpen
  \bibfield  {author} {\bibinfo {author} {\bibfnamefont {Y.~V.}\ \bibnamefont
  {Kartashov}}, \bibinfo {author} {\bibfnamefont {B.~A.}\ \bibnamefont
  {Malomed}}, \ and\ \bibinfo {author} {\bibfnamefont {L.}~\bibnamefont
  {Torner}},\ }\href {\doibase 10.1103/PhysRevLett.122.193902} {\bibfield
  {journal} {\bibinfo  {journal} {Phys. Rev. Lett.}\ }\textbf {\bibinfo
  {volume} {122}},\ \bibinfo {pages} {193902} (\bibinfo {year}
  {2019})}\BibitemShut {NoStop}%
\bibitem [{\citenamefont {Cidrim}\ \emph {et~al.}(2018)\citenamefont {Cidrim},
  \citenamefont {dos Santos}, \citenamefont {Henn},\ and\ \citenamefont
  {Macr\`{\i}}}]{cidrim2018}%
  \BibitemOpen
  \bibfield  {author} {\bibinfo {author} {\bibfnamefont {A.}~\bibnamefont
  {Cidrim}}, \bibinfo {author} {\bibfnamefont {F.~E.~A.}\ \bibnamefont {dos
  Santos}}, \bibinfo {author} {\bibfnamefont {E.~A.~L.}\ \bibnamefont {Henn}},
  \ and\ \bibinfo {author} {\bibfnamefont {T.}~\bibnamefont {Macr\`{\i}}},\
  }\href {\doibase 10.1103/PhysRevA.98.023618} {\bibfield  {journal} {\bibinfo
  {journal} {Phys. Rev. A}\ }\textbf {\bibinfo {volume} {98}},\ \bibinfo
  {pages} {023618} (\bibinfo {year} {2018})}\BibitemShut {NoStop}%
\bibitem [{\citenamefont {Li}\ \emph {et~al.}(2018)\citenamefont {Li},
  \citenamefont {Chen}, \citenamefont {Luo}, \citenamefont {Huang},
  \citenamefont {Tan}, \citenamefont {Pang},\ and\ \citenamefont
  {Malomed}}]{li2018}%
  \BibitemOpen
  \bibfield  {author} {\bibinfo {author} {\bibfnamefont {Y.}~\bibnamefont
  {Li}}, \bibinfo {author} {\bibfnamefont {Z.}~\bibnamefont {Chen}}, \bibinfo
  {author} {\bibfnamefont {Z.}~\bibnamefont {Luo}}, \bibinfo {author}
  {\bibfnamefont {C.}~\bibnamefont {Huang}}, \bibinfo {author} {\bibfnamefont
  {H.}~\bibnamefont {Tan}}, \bibinfo {author} {\bibfnamefont {W.}~\bibnamefont
  {Pang}}, \ and\ \bibinfo {author} {\bibfnamefont {B.~A.}\ \bibnamefont
  {Malomed}},\ }\href {\doibase 10.1103/PhysRevA.98.063602} {\bibfield
  {journal} {\bibinfo  {journal} {Phys. Rev. A}\ }\textbf {\bibinfo {volume}
  {98}},\ \bibinfo {pages} {063602} (\bibinfo {year} {2018})}\BibitemShut
  {NoStop}%
\bibitem [{\citenamefont {Kartashov}\ \emph {et~al.}(2018)\citenamefont
  {Kartashov}, \citenamefont {Malomed}, \citenamefont {Tarruell},\ and\
  \citenamefont {Torner}}]{kartashov2018}%
  \BibitemOpen
  \bibfield  {author} {\bibinfo {author} {\bibfnamefont {Y.~V.}\ \bibnamefont
  {Kartashov}}, \bibinfo {author} {\bibfnamefont {B.~A.}\ \bibnamefont
  {Malomed}}, \bibinfo {author} {\bibfnamefont {L.}~\bibnamefont {Tarruell}}, \
  and\ \bibinfo {author} {\bibfnamefont {L.}~\bibnamefont {Torner}},\ }\href
  {\doibase 10.1103/PhysRevA.98.013612} {\bibfield  {journal} {\bibinfo
  {journal} {Phys. Rev. A}\ }\textbf {\bibinfo {volume} {98}},\ \bibinfo
  {pages} {013612} (\bibinfo {year} {2018})}\BibitemShut {NoStop}%
\bibitem [{\citenamefont {Ancilotto}\ \emph
  {et~al.}(2018{\natexlab{b}})\citenamefont {Ancilotto}, \citenamefont
  {Barranco}, \citenamefont {Guilleumas},\ and\ \citenamefont
  {Pi}}]{ancilotto2018b}%
  \BibitemOpen
  \bibfield  {author} {\bibinfo {author} {\bibfnamefont {F.}~\bibnamefont
  {Ancilotto}}, \bibinfo {author} {\bibfnamefont {M.}~\bibnamefont {Barranco}},
  \bibinfo {author} {\bibfnamefont {M.}~\bibnamefont {Guilleumas}}, \ and\
  \bibinfo {author} {\bibfnamefont {M.}~\bibnamefont {Pi}},\ }\href {\doibase
  10.1103/PhysRevA.98.053623} {\bibfield  {journal} {\bibinfo  {journal} {Phys.
  Rev. A}\ }\textbf {\bibinfo {volume} {98}},\ \bibinfo {pages} {053623}
  (\bibinfo {year} {2018}{\natexlab{b}})}\BibitemShut {NoStop}%
\bibitem [{\citenamefont {Chin}\ and\ \citenamefont
  {Krotscheck}(2005)}]{chinkrotscheck2005}%
  \BibitemOpen
  \bibfield  {author} {\bibinfo {author} {\bibfnamefont {S.~A.}\ \bibnamefont
  {Chin}}\ and\ \bibinfo {author} {\bibfnamefont {E.}~\bibnamefont
  {Krotscheck}},\ }\href {\doibase 10.1103/PhysRevE.72.036705} {\bibfield
  {journal} {\bibinfo  {journal} {Phys. Rev. E}\ }\textbf {\bibinfo {volume}
  {72}},\ \bibinfo {pages} {036705} (\bibinfo {year} {2005})}\BibitemShut
  {NoStop}%
\bibitem [{\citenamefont {Komineas}\ \emph {et~al.}(2005)\citenamefont
  {Komineas}, \citenamefont {Cooper},\ and\ \citenamefont
  {Papanicolaou}}]{komineas2005}%
  \BibitemOpen
  \bibfield  {author} {\bibinfo {author} {\bibfnamefont {S.}~\bibnamefont
  {Komineas}}, \bibinfo {author} {\bibfnamefont {N.~R.}\ \bibnamefont
  {Cooper}}, \ and\ \bibinfo {author} {\bibfnamefont {N.}~\bibnamefont
  {Papanicolaou}},\ }\href {\doibase 10.1103/PhysRevA.72.053624} {\bibfield
  {journal} {\bibinfo  {journal} {Phys. Rev. A}\ }\textbf {\bibinfo {volume}
  {72}},\ \bibinfo {pages} {053624} (\bibinfo {year} {2005})}\BibitemShut
  {NoStop}%
\bibitem [{\citenamefont {Bretin}\ \emph {et~al.}(2004)\citenamefont {Bretin},
  \citenamefont {Stock}, \citenamefont {Seurin},\ and\ \citenamefont
  {Dalibard}}]{bretin2004}%
  \BibitemOpen
  \bibfield  {author} {\bibinfo {author} {\bibfnamefont {V.}~\bibnamefont
  {Bretin}}, \bibinfo {author} {\bibfnamefont {S.}~\bibnamefont {Stock}},
  \bibinfo {author} {\bibfnamefont {Y.}~\bibnamefont {Seurin}}, \ and\ \bibinfo
  {author} {\bibfnamefont {J.}~\bibnamefont {Dalibard}},\ }\href {\doibase
  10.1103/PhysRevLett.92.050403} {\bibfield  {journal} {\bibinfo  {journal}
  {Phys. Rev. Lett.}\ }\textbf {\bibinfo {volume} {92}},\ \bibinfo {pages}
  {050403} (\bibinfo {year} {2004})}\BibitemShut {NoStop}%
\bibitem [{\citenamefont {K\"arkk\"ainen}\ \emph {et~al.}(2007)\citenamefont
  {K\"arkk\"ainen}, \citenamefont {Christensson}, \citenamefont {Reinisch},
  \citenamefont {Kavoulakis},\ and\ \citenamefont {Reimann}}]{karkkainen2007}%
  \BibitemOpen
  \bibfield  {author} {\bibinfo {author} {\bibfnamefont {K.}~\bibnamefont
  {K\"arkk\"ainen}}, \bibinfo {author} {\bibfnamefont {J.}~\bibnamefont
  {Christensson}}, \bibinfo {author} {\bibfnamefont {G.}~\bibnamefont
  {Reinisch}}, \bibinfo {author} {\bibfnamefont {G.~M.}\ \bibnamefont
  {Kavoulakis}}, \ and\ \bibinfo {author} {\bibfnamefont {S.~M.}\ \bibnamefont
  {Reimann}},\ }\href {\doibase 10.1103/PhysRevA.76.043627} {\bibfield
  {journal} {\bibinfo  {journal} {Phys. Rev. A}\ }\textbf {\bibinfo {volume}
  {76}},\ \bibinfo {pages} {043627} (\bibinfo {year} {2007})}\BibitemShut
  {NoStop}%
\bibitem [{\citenamefont {Petrov}\ and\ \citenamefont
  {Shlyapnikov}(2001)}]{petrov2001}%
  \BibitemOpen
  \bibfield  {author} {\bibinfo {author} {\bibfnamefont {D.~S.}\ \bibnamefont
  {Petrov}}\ and\ \bibinfo {author} {\bibfnamefont {G.~V.}\ \bibnamefont
  {Shlyapnikov}},\ }\href {\doibase 10.1103/PhysRevA.64.012706} {\bibfield
  {journal} {\bibinfo  {journal} {Phys. Rev. A}\ }\textbf {\bibinfo {volume}
  {64}},\ \bibinfo {pages} {012706} (\bibinfo {year} {2001})}\BibitemShut
  {NoStop}%
\bibitem [{\citenamefont {Ilg}\ \emph {et~al.}(2018)\citenamefont {Ilg},
  \citenamefont {Kumlin}, \citenamefont {Santos}, \citenamefont {Petrov},\ and\
  \citenamefont {B\"uchler}}]{ilg2018}%
  \BibitemOpen
  \bibfield  {author} {\bibinfo {author} {\bibfnamefont {T.}~\bibnamefont
  {Ilg}}, \bibinfo {author} {\bibfnamefont {J.}~\bibnamefont {Kumlin}},
  \bibinfo {author} {\bibfnamefont {L.}~\bibnamefont {Santos}}, \bibinfo
  {author} {\bibfnamefont {D.~S.}\ \bibnamefont {Petrov}}, \ and\ \bibinfo
  {author} {\bibfnamefont {H.~P.}\ \bibnamefont {B\"uchler}},\ }\href {\doibase
  10.1103/PhysRevA.98.051604} {\bibfield  {journal} {\bibinfo  {journal} {Phys.
  Rev. A}\ }\textbf {\bibinfo {volume} {98}},\ \bibinfo {pages} {051604}
  (\bibinfo {year} {2018})}\BibitemShut {NoStop}%
\end{thebibliography}%

\end{document}